\documentclass[%
reprint,
superscriptaddress,
preprintnumbers,
nobibnotes,
amsmath,amssymb,
aps,
prc,
]{revtex4-2}
\usepackage[pdftex]{graphicx}
\usepackage{dcolumn}
\usepackage{bm}
\usepackage[pdftex]{color}
\usepackage{CJK}
\usepackage[T1]{fontenc}
\usepackage{mathrsfs}
\def\ve#1{{\bm{#1}}}
\def\nuc#1#2#3{{}^{#2}_{#3}\mathrm{#1}}
\def\urm#1{\scriptstyle{\text{\textrm{\textmd{\textup{#1}}}}}}

\let\temp\epsilon
\let\epsilon\varepsilon
\let\varepsilon\temp
\let\temp\relax
\let\temp\phi
\let\phi\varphi
\let\varphi\temp
\let\temp\relax
\begin{document}
%
\begin{CJK*}{UTF8}{}
  \preprint{RIKEN-iTHEMS-Report-24}
  \title{$ {}^{164} \mathrm{Pb} $: A possible heaviest $ N = Z $ doubly magic nucleus}
  \author{Tomoya Naito (\CJKfamily{min}{内藤智也})}
  \email{
    tnaito@ribf.riken.jp}
  \affiliation{
    RIKEN Interdisciplinary Theoretical and Mathematical Sciences Program (iTHEMS),
    Wako 351-0198, Japan}
  \affiliation{
    Department of Physics, Graduate School of Science, The University of Tokyo,
    Tokyo 113-0033, Japan}
  \author{Masaaki Kimura (\CJKfamily{min}{木村真明})}
  \email{
    masaaki.kimura@ribf.riken.jp}
  \affiliation{
    RIKEN Nishina Center for Accelerator-Based Science,
    Wako 351-0198, Japan}
  \affiliation{
    Department of Physics, Graduate School of Science, The University of Tokyo,
    Tokyo 113-0033, Japan}
  \author{Masaki Sasano (\CJKfamily{min}{笹野匡紀})}
  \email{
    sasano@ribf.riken.jp}
  \affiliation{
    RIKEN Nishina Center for Accelerator-Based Science,
    Wako 351-0198, Japan}
  \date{\today}
  \begin{abstract}
    We confirm by using the Skyrme Hartree-Fock-Bogoliubov calculation that
    $ {}^{164} \mathrm{Pb} $ is a possible heaviest $ N = Z $ doubly magic nucleus
    whose lifetime is long enough to be measured on accelerator experiments.
    We estimate the proton-emission and alpha-decay half-lives of $ {}^{164} \mathrm{Pb} $.
    The estimated proton-emission half-life ranges from $ 0.1 \, \mathrm{ps} $ to $ 10 \, \mathrm{ns} $,
    while the alpha decay can be safely neglected.
  \end{abstract}
  \maketitle
\end{CJK*}
%
\section{Introduction}
\par
With the advent of accelerator facilities worldwide, the nuclear chart has been rapidly extended toward the regions where the neutron or proton number far exceeds that of naturally abundant nuclei~\cite{
  Nakamura2017Prog.Part.Nucl.Phys.97_53,            
  Sakurai2018Front.Phys.13_132111}.
A fundamental question motivating such experimental challenges is
``How far from the beta stability line can atomic nuclei exist with a finite half-life?,''
which serves as a rigorous test for our understanding of the nuclear binding~\cite{
  Casten2000Prog.Part.Nucl.Phys.45_S171,
  Paar2010J.Phys.G37_064014,
  Zhou2017Proc.Sci.281_373,
  Nakada2020Int.J.Mod.Phys.E29_1930008}.
\par
For each isotopic chain, the binding energy of the outermost valence neutron defines the limit of neutron excess. 
In contrast, the limit of proton excess remains ambiguous both experimentally and theoretically. 
This ambiguity arises because, in extremely proton-rich nuclei,
 proton orbitals with positive energy can be meta-stable with the assistance of the Coulomb barrier, 
allowing them to exist with finite half-lives as non-Hermitian/open systems that undergo charged-particle emission. 
This concept dates back to the early works of 
Gamow~\cite{
  Gamow1934Z.Phys.89_592},
Sigert~\cite{
  Siegert1939Phys.Rev.56_750}, 
Majorana~\cite{
  Majorana2006Electron.J.Theor.Phys.3_293}, 
and Feshbach~\cite{
  Feshbach1958Ann.Phys.5_357,
  Feshbach1962Ann.Phys.19_287}.
Consequently, the exploration of extremely proton-rich nuclei illuminates understanding of open quantum systems, a topic of interest across various disciplines of physics~\cite{
  Szankowski2023SciPostPhys.Lect.Notes_68}. 
In addition, such extreme systems are expected to play an important role in understanding the symmetry energy of nuclear matter, as well as the isospin symmetry of nuclear forces and its breaking~\cite{
  Heisenberg1932Z.Phys.77_1,
  Cassen1936Phys.Rev.50_846,
  Wigner1937Phys.Rev.51_106,
  Auerbach1969Phys.Rev.Lett.23_484,
  Shlomo1978Rep.Prog.Phys.41_957,
  Baczyk2018Phys.Lett.B778_178,
  Baczyk2019J.Phys.G46_03LT01,
  Sagawa2019Eur.Phys.J.A55_227,
  Naito2022Phys.Rev.C105_L021304,
  Naito2022Phys.Rev.C106_L061306,
  Sagawa2022Phys.Lett.B829_137072,
  Naito2023Phys.Rev.C107_064302,
  Sagawa2024Phys.Rev.C109_L011302}.
\par
The decay lifetime of states confined by the Coulomb barrier depends exponentially on the difference between the Coulomb barrier height and the binding energy. 
If a shell gap exists in the proton orbitals, the decay lifetime should differ by orders of magnitude above and below the gap. 
Therefore, a doubly closed-shell nucleus likely represents the limit of the existence of proton-rich nuclei. 
In this sense, $ \nuc{Pb}{164}{} $ is a candidate for the last doubly closed-shell nucleus following $ \nuc{Sn}{100}{} $,
and it could be the ultimate endpoint for proton-rich nuclei.
However, it is not trivial whether the magic numbers remain intact in a mass region significantly different from ordinary nuclei. 
For instance, the breakdown of the standard shell structure has been predicted for superheavy nuclei~\cite{
  Kruppa2000Phys.Rev.C61_034313,
  Sobiczewski2007Prog.Part.Nucl.Phys.58_292,
  Li2014Phys.Lett.B732_169,
  Giuliani2019Rev.Mod.Phys.91_011001}.
In particular, one may expect that the strong Coulomb repulsion affects the order of single-particle levels.
\par
Furthermore, it is also emphasized that $ \nuc{Pb}{164}{} $ has the potential to serve as an excellent benchmark for energy density functionals (EDFs),
particularly in terms of isospin symmetry breaking.
The proton-neutron asymmetry in $ N = Z $ nuclei,
such as the proton-skin thickness~\cite{
  Naito2023Phys.Rev.C107_064302}
and the isovector density~\cite{
  Sagawa2022Phys.Lett.B829_137072},
can be used to extract information about the breaking of isospin symmetry in atomic nuclei.
In this regard, $ \nuc{Pb}{164}{} $ is an extreme $ N = Z $ nucleus;
therefore, its properties should be instrumental in gaining a deeper understanding of isospin symmetry breaking in atomic nuclei.
\par
Therefore, in this paper, to seek the possibility of a finite half-life of $ \nuc{Pb}{164}{} $ and whether $ Z = N = 82 $ magicity holds,
we investigate this nucleus 
using the Skyrme Hartree-Fock-Bogoliubov (HFB) calculation~\cite{
  Vautherin1972Phys.Rev.C5_626}
with continuum.
Since some protons occupy near or above the threshold, it is crucial to consider the pairing correlation coupling to the continuum.
Then, the half-life of the one-proton emission is estimated by using the Wentzel-Kramers-Brillouin (WKB) approximation~\cite{
  Buck1992Phys.Rev.C45_1688,
  Aaberg1997Phys.Rev.C56_1762,
  Xiao2023Phys.Lett.B845_138160}.
The alpha-decay half-life is also estimated by using a phenomenological formula~\cite{
  Koura2012J.Nucl.Sci.Technol.49_816}.
%
\section{Calculation setup}
\par
We perform the Skyrme HFB calculations~\cite{
  Vautherin1972Phys.Rev.C5_626,
  Dobaczewski1984Nucl.Phys.A422_103}
with the typical six EDFs---SLy4~\cite{
  Chabanat1998Nucl.Phys.A635_231}, 
SLy5~\cite{
  Chabanat1998Nucl.Phys.A635_231},
SkM*~\cite{
  Bartel1982Nucl.Phys.A386_79},
UNEDF0~\cite{
  Kortelainen2010Phys.Rev.C82_024313},
UNEDF1~\cite{
  Kortelainen2012Phys.Rev.C85_024304},
and UNEDF2~\cite{
  Kortelainen2014Phys.Rev.C89_054314}.
The volume-type pairing interaction is used,
where the cutoff energy is $ 60 \, \mathrm{MeV} $ in the Hartree-Fock equivalent energy~\cite{
  NavarroPerez2017Comput.Phys.Commun.220_363}
and the pairing strength is determined to reproduce the neutron pairing gap of $ \nuc{Sn}{120}{} $ as $ 1.4 \, \mathrm{MeV} $~\cite{
  Naito2023Phys.Rev.C107_054307}.
The spherical symmetry is assumed in the calculation~\footnote{
  In Appendix~\ref{sec:appendix_HFBTHO},
  the Skyrme HFB calculation with considering the axial deformation
  using the harmonic-oscillator basis 
  is performed using the HFBTHO code~\cite{
    NavarroPerez2017Comput.Phys.Commun.220_363}.
  Actually, it was confirmed that $ \nuc{Pb}{164}{} $ is spherical and shows doubly magic properties
  even if the axial deformation is considered.}.
The radial coordinate is discretized and
the HFB eigenvalue is obtained by diagonalizing the single-particle HFB Hamiltonian.
Consequently, the pairing correlation with continuum is considered automatically.
\par
We calculated with $ 160 \times 0.1 \, \mathrm{fm} $ and $ 240 \times 0.1 \, \mathrm{fm} $ box sizes,
and found that the $ 160 \times 0.1 \, \mathrm{fm} $ box yields converged results.
We also performed the Hartree-Fock calculation with a $ 480 \times 0.1 \, \mathrm{fm} $ box to evaluate the proton-decay half-life of $ \nuc{Pb}{164}{} $ with the WKB approximation.
%
\section{Calculation results}
\subsection{Binding energy and magicity}
\par
In order to calculate the two-proton separation energy $ S_{2p} $ and the $ \alpha $-decay $ Q $-value $ Q_{\alpha} $,
we also calculate $ \nuc{Hg}{160}{} $ and $ \nuc{Hg}{162}{} $, as well as $ \nuc{Pb}{164}{} $.
Once the energies of these nuclei $ E $
are obtained, 
$ S_{2p} $ and $ Q_{\alpha} $ of $ \nuc{Pb}{164}{} $ are, respectively, calculated by
\begin{subequations}
  \begin{align}
    S_{2p} 
    & =
      E \left( \nuc{Hg}{162}{} \right)
      -
      E \left( \nuc{Pb}{164}{} \right), \\
    Q_{\alpha} 
    & =
      E \left( \nuc{Pb}{164}{} \right)
      -
      E \left( \nuc{Hg}{160}{} \right)
      -
      E \left( \nuc{He}{4}{} \right).
  \end{align}
\end{subequations}
The experimental value 
$ 28.2957 \, \mathrm{MeV} $~\cite{
  Huang2021Chin.Phys.C45_030002,
  Wang2021Chin.Phys.C45_030003}
is used for $ E \left( \nuc{He}{4}{} \right) $.
\par
The one-proton separation energy $ S_p $ is defined by
\begin{equation}
  \label{eq:Sp}
  S_p
  =
  E \left( \nuc{Tl}{163}{} \right)
  -
  E \left( \nuc{Pb}{164}{} \right).
\end{equation}
In this work, the energy of $ \nuc{Tl}{163}{} $ is estimated in an approximated way;
the three-point formula for the pairing gap for $ \nuc{Tl}{163}{} $ is given by~\cite{
  Bender2000Eur.Phys.J.A8_59}
\begin{equation}
  \label{eq:3p_gap}
  \Delta_p \left( \nuc{Tl}{163}{} \right)
  \simeq
  -
  \frac{E \left( \nuc{Hg}{162}{} \right) - 2 E \left( \nuc{Tl}{163}{} \right) + E \left( \nuc{Pb}{164}{} \right)}{2}.
\end{equation}
Consequently, $ E \left( \nuc{Tl}{163}{} \right) $ reads
\begin{equation}
  E \left( \nuc{Tl}{163}{} \right)
  \simeq
  \frac{E \left( \nuc{Hg}{162}{} \right) + E \left( \nuc{Pb}{164}{} \right)}{2}
  +
  \Delta_p \left( \nuc{Tl}{163}{} \right).
\end{equation}
According to Ref.~\cite{
  Bender2000Eur.Phys.J.A8_59},
$ \Delta_p $ is a smooth function except at the magic numbers;
hence, we take an additional approximation
$ \Delta_p \left( \nuc{Tl}{163}{} \right) \simeq \Delta_p \left( \nuc{Hg}{162}{} \right) $,
and accordingly, $ E \left( \nuc{Tl}{163}{} \right) $ is approximated by
\begin{equation}
  \label{eq:Sp_approx}
  E \left( \nuc{Tl}{163}{} \right)
  \simeq
  \frac{E \left( \nuc{Hg}{162}{} \right) + E \left( \nuc{Pb}{164}{} \right)}{2}
  +
  \Delta_p \left( \nuc{Hg}{162}{} \right).
\end{equation}
The pairing gap $ \Delta_p $ is estimated by the average gap~\cite{
  Bender2000Eur.Phys.J.A8_59}
\begin{equation}
  \Delta_p
  =
  \frac{\int \tilde{V}_p \left( \ve{r} \right) \rho_p \left( \ve{r} \right) \, d \ve{r}}{\int \rho_p \left( \ve{r} \right) \, d \ve{r}}
  =
  \frac{\int \tilde{V}_p \left( \ve{r} \right) \rho_p \left( \ve{r} \right) \, d \ve{r}}{Z},
\end{equation}
where $ \tilde{V}_p $ is the pairing mean-field potential for protons
and $ \rho_p $ is the proton density~\footnote{
  Another way to calculate the pairing gap $ \Delta_p \left( \nuc{Hg}{162}{} \right) $ is the spectral gap
  \begin{equation}
    \tilde{\Delta}_p
    =
    \frac{\int \tilde{V}_p \left( \ve{r} \right) \tilde{\rho}_p \left( \ve{r} \right) \, d \ve{r}}{\int \tilde{\rho}_p \left( \ve{r} \right) \, d \ve{r}}
  \end{equation}
  where $ \tilde{\rho}_p $ is the proton pair density.
  The results do not change much even if we use the spectral gap instead.}.
\par
In order to examine which EDFs are most appropriate,
the same prescription is applied to $ \nuc{Sn}{100}{} $,
whose results are shown in Appendix~\ref{sec:appendix_Sn100}.
The difference between theories and experiment in $ S_{2p} $ and $ Q_{\alpha} $ are at most $ 1 $ and $ 3.5 \, \mathrm{MeV} $, respectively.
The accuracy of $ S_p $ also depends on the approximation given in Eq.~\eqref{eq:Sp_general},
which is the generalized version of Eq.~\eqref{eq:Sp_approx}.
The theoretical values of $ S_p $ are within $ 1 \, \mathrm{MeV} $ difference from the experimental data,
which validates the approximation given in Eqs.~\eqref{eq:Sp_approx} and \eqref{eq:Sp_general}.
Thus, these six EDFs give reasonable results,
while SLy4, SLy5, and UNEDF0 are more accurate.
Therefore, hereinafter, we mainly refer to the results obtained by the SLy4.
\par
\begin{figure*}[tb]
  \centering
  \includegraphics[width=1.0\linewidth]{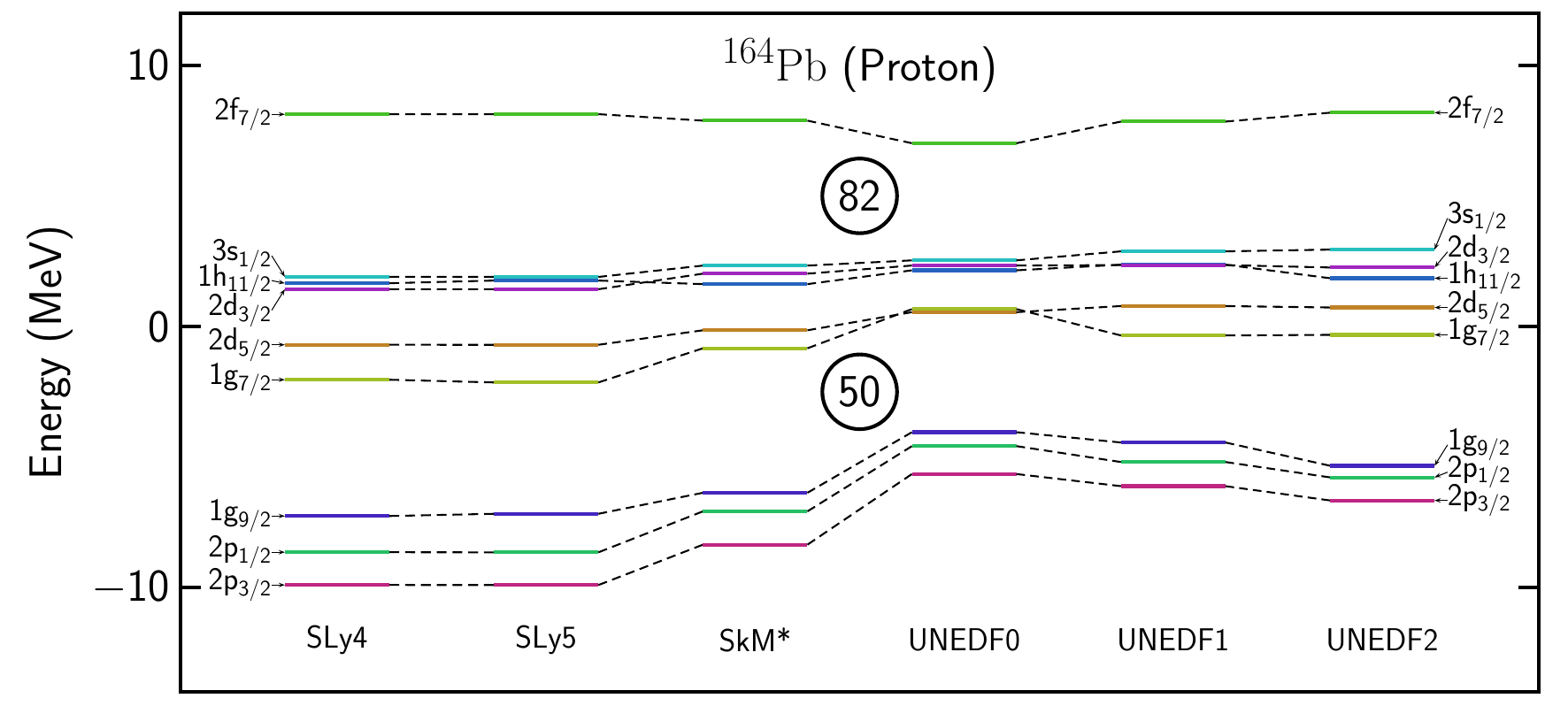}
  \caption{
    Proton single-particle spectra of $ \nuc{Pb}{164}{} $
    calculated by using the Skyrme Hartree-Fock-Bogoliubov calculation with a $ 160 \times 0.1 \, \mathrm{fm} $ box.
    The lowest unoccupied state, the $ 2f_{7/2} $ orbital, is also shown.
    The single-particle energies plotted here are eigenvalues of the $ p $-$ h $ Hamiltonian.}
  \label{fig:Pb164_sp}
\end{figure*}

Table~\ref{tab:Pb164_result} shows the results for $ \nuc{Pb}{164}{} $
and Fig.~\ref{fig:Pb164_sp} shows proton single-particle spectra.
The proton lowest unoccupied orbital is the $ 2f_{7/2} $,
whose energy is approximately $ 8 \, \mathrm{MeV} $,
and the pairing gap of $ \nuc{Pb}{164}{} $ is less than $ 10 \, \mathrm{keV} $.
Hence, the $ Z = 82 $ shell gap is large enough.
Furthermore, in Appendix~\ref{sec:appendix_HFBTHO},
the $ N = Z $ nuclei are calculated within the axial deformation~\cite{
  NavarroPerez2017Comput.Phys.Commun.220_363}.
The systematics of $ \Delta $, $ S_{2p} $, $ Q_{\alpha} $, and the deformation parameter $ \beta_2 $ also show
that $ \nuc{Pb}{164}{} $ is a typical doubly magic nucleus.
Therefore, we conclude that $ Z = N = 82 $ magicity still holds.
\par
\begin{figure}[tb]
  \centering
  \includegraphics[width=1.0\linewidth]{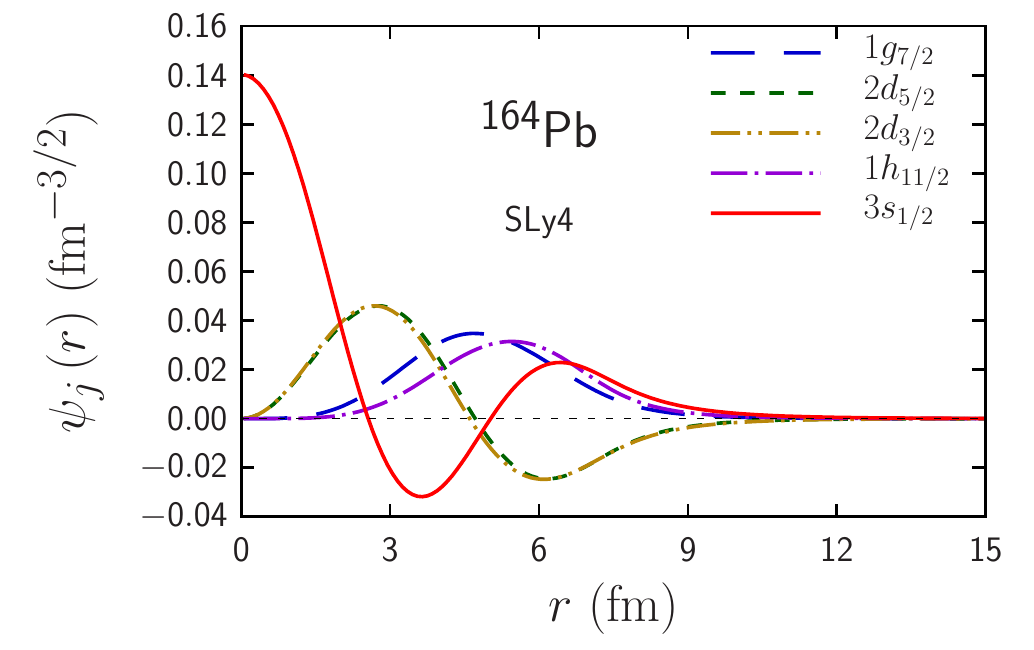}
  \caption{
    Proton single-particle orbitals of $ \nuc{Pb}{164}{} $
    calculated by using the Skyrme Hartree-Fock calculation with a $ 480 \times 0.1 \, \mathrm{fm} $ box
    with the SLy4 EDF.}
  \label{fig:Pb164_sp_wf}
\end{figure}
\begin{figure}[tb]
  \centering
  \includegraphics[width=1.0\linewidth]{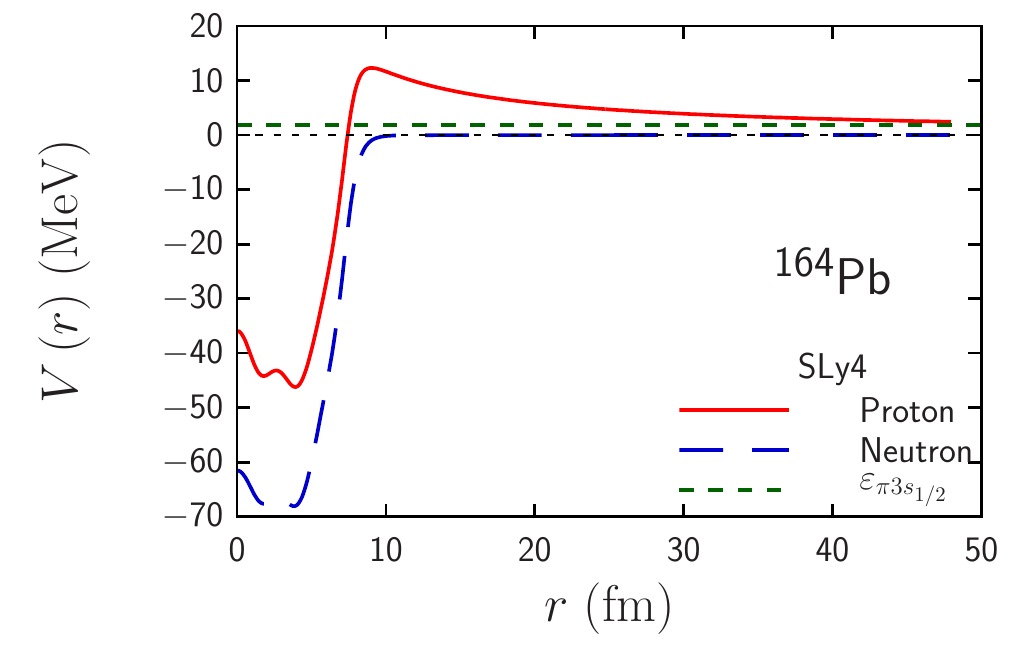}
  \caption{
    Proton and neutron mean-field potentials of $ \nuc{Pb}{164}{} $
    calculated by using the Skyrme Hartree-Fock calculation with a $ 480 \times 0.1 \, \mathrm{fm} $ box
    with the SLy4 EDF.
    The single-particle energy of the proton $ 3s_{1/2} $ orbital,
    $ \epsilon_{\pi 3s_{1/2}} $, is shown in the green dotted line.}
  \label{fig:Pb164_pot}
\end{figure}
\begin{figure}[tb]
  \centering
  \includegraphics[width=1.0\linewidth]{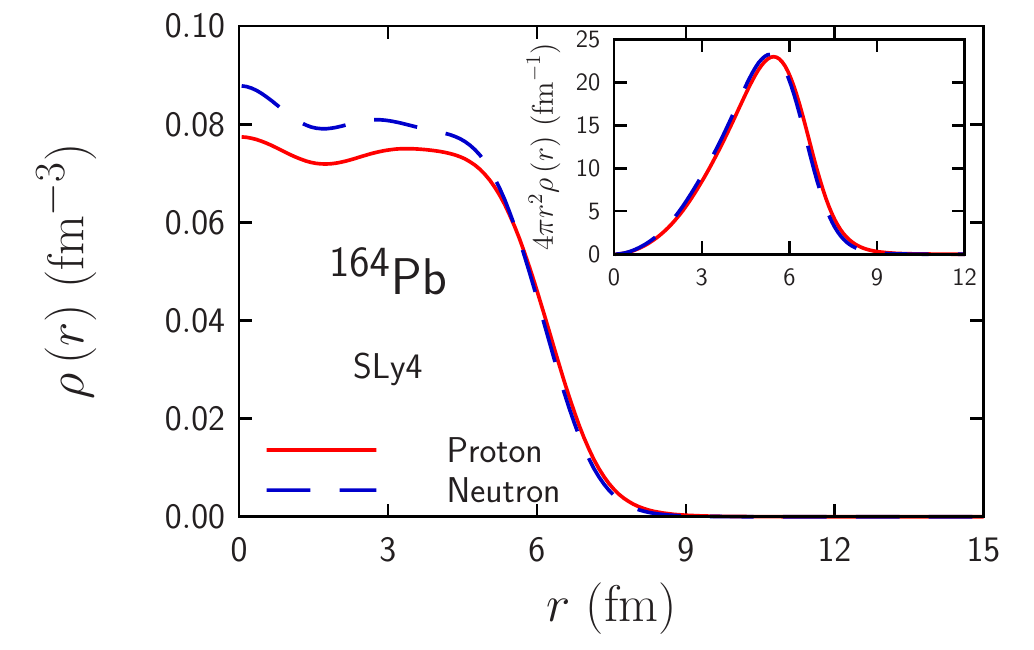}
  \caption{
    Proton and neutron densities of $ \nuc{Pb}{164}{} $
    calculated by using the Skyrme Hartree-Fock calculation with $ 480 \times 0.1 \, \mathrm{fm} $ box
    with the SLy4 EDF.}
  \label{fig:Pb164_dens}
\end{figure}

Despite considerable neutron deficiency,
single-particle spectra are still similar to those of stable isotopes
as shown in Fig.~\ref{fig:Pb164_sp}.
The proton highest occupied single-particle orbital is the $ 3s_{1/2} $,
which is not a standard shell structure~\cite{
  Cwiok1996Nucl.Phys.A611_211,
  Naito2023Phys.Rev.C107_054307}.
The proton single-particle orbitals of the $ sdg $ shell shown in Fig.~\ref{fig:Pb164_sp_wf}
behave similarly to the bound orbitals
and are localized in $ r \lesssim 10 \, \mathrm{fm} $,
inside the Coulomb barrier.
The bottom of the proton mean field is about $ 20 \, \mathrm{MeV} $ higher than the neutron one
(see~Fig.~\ref{fig:Pb164_pot}).
Accordingly, the neutron single-particle energies are also about $ 20 \, \mathrm{MeV} $ deeper.
This leads to the large asymmetry in the proton-neutron distribution at the central region (see~Fig.~\ref{fig:Pb164_dens}).
\par
The density distribution of $ \nuc{Pb}{164}{} $ is shown in Fig.~\ref{fig:Pb164_dens}.
The proton and neutron root-mean-square radii, $ R_p $ and $ R_n $,
and
the charge radii $ R_{\urm{ch}} $ calculated by using Eq.~(2) of Ref.~\cite{
  Naito2023Phys.Rev.C107_054307}
are summarized in Table~\ref{tab:Pb164_result}.
Because of the strong Coulomb repulsion,
$ R_p $ is much larger than $ R_n $
and the proton skin $ R_p - R_n $ reaches $ 0.1 \, \mathrm{fm} $.
The peak position of $ 4 \pi r^2 \rho_p $ is also about $ 0.1 \, \mathrm{fm} $ outside of $ 4 \pi r^2 \rho_n $,
which is consistent with the proton-skin thickness.
\par
\begin{figure}[tb]
  \centering
  \includegraphics[width=1.0\linewidth]{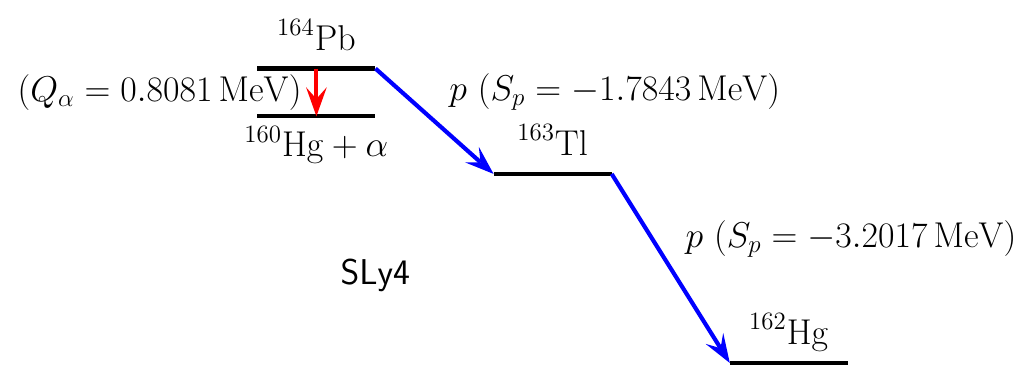}
  \caption{
    Decay scheme of $ \nuc{Pb}{164}{} $
    calculated by using the Skyrme Hartree-Fock-Bogoliubov calculation with a $ 160 \times 0.1 \, \mathrm{fm} $ box
    with the SLy4 EDF.}
  \label{fig:Pb164_decay_scheme}
\end{figure}

As all the EDFs yield $ S_p < 0 $, $ S_{2p} < 0 $, and $ Q_{\alpha} > 0 $,
one- and two-proton emission, and the $ \alpha $ decay channels should be opened
as summarized in Fig.~\ref{fig:Pb164_decay_scheme}.
Since the proton highest single-particle orbital is the $ 3s_{1/2} $,
the $ 1p $ emission from this orbital should dominate, and
the $ 2p $-emission process is negligible.
\begin{table*}[tb]
  \centering
  \caption{
    Calculation results for $ \nuc{Pb}{164}{} $.
    These calculation results are obtained by the Skyrme Hartree-Fock-Bogoloubov calculation
    with the
    SLy4~\cite{
      Chabanat1998Nucl.Phys.A635_231}, 
    SLy5~\cite{
      Chabanat1998Nucl.Phys.A635_231},
    SkM*~\cite{
      Bartel1982Nucl.Phys.A386_79},
    UNEDF0~\cite{
      Kortelainen2010Phys.Rev.C82_024313},
    UNEDF1~\cite{
      Kortelainen2012Phys.Rev.C85_024304},
    and UNEDF2~\cite{
      Kortelainen2014Phys.Rev.C89_054314}
    EDFs
    and the volume-type pairing interaction.
    All the results except $ R_p $, $ R_n $, or $ R_{\urm{ch}} $ are shown in $ \mathrm{MeV} $,
    while $ R_p $, $ R_n $, are $ R_{\urm{ch}} $ are shown in $ \mathrm{fm} $.}
  \label{tab:Pb164_result}
  \begin{ruledtabular}
    \begin{tabular}{ldddddd}
      EDF & \multicolumn{1}{c}{SLy4} & \multicolumn{1}{c}{SLy5} & \multicolumn{1}{c}{SkM*} & \multicolumn{1}{c}{UNEDF0} & \multicolumn{1}{c}{UNEDF1} & \multicolumn{1}{c}{UNEDF2} \\
      \hline
      $ E \left( \nuc{Pb}{164}{} \right) $        & -1208.3863 & -1207.4446 & -1198.3397 & -1207.5137 & -1210.0435 & -1210.2978 \\
      $ E \left( \nuc{Hg}{162}{} \right) $        & -1213.3723 & -1212.4313 & -1203.2334 & -1213.0773 & -1215.8523 & -1215.7930 \\
      $ E \left( \nuc{Hg}{160}{} \right) $        & -1180.8988 & -1180.0439 & -1170.4134 & -1181.5332 & -1184.0256 & -1183.3580 \\
      $ E \left( \nuc{Tl}{163}{} \right) $        & -1210.1707 & -1209.2717 & -1200.2815 & -1209.7663 & -1212.3952 & -1212.4658 \\
      \hline
      $ S_p $                                     & -1.7843 & -1.8271 & -1.9419 & -2.2525 & -2.3517 & -2.1680 \\
      $ S_{2p} $                                  & -4.9860 & -4.9867 & -4.8937 & -5.5636 & -5.8088 & -5.4952 \\
      $ Q_{\alpha} $                              & 0.8081 & 0.8950 & 0.3694 & 2.3151 & 2.2778 & 1.3558 \\
      \hline
      $ \Delta_p \left( \nuc{Pb}{164}{} \right) $ & 0.0006 & 0.0005 & 0.0003 & 0.0004 & 0.0005 & 0.0005 \\
      $ \Delta_n \left( \nuc{Pb}{164}{} \right) $ & 0.0006 & 0.0004 & 0.0000 & 0.0013 & 0.0009 & 0.0007 \\
      $ \Delta_p \left( \nuc{Hg}{162}{} \right) $ & 0.7086 & 0.6662 & 0.5050 & 0.5293 & 0.5528 & 0.5796 \\
      $ \Delta_n \left( \nuc{Hg}{162}{} \right) $ & 0.0022 & 0.0018 & 0.0009 & 0.0031 & 0.0027 & 0.0020 \\
      \hline
      $ \epsilon_{\pi 1g_{7/2}} $                 & -2.0381 & -2.1416 & -0.8348 & +0.6652 & -0.3438 & -0.3199 \\
      $ \epsilon_{\pi 2d_{5/2}} $                 & -0.6997 & -0.7006 & -0.1450 & +0.5489 & +0.7821 & +0.7275 \\
      $ \epsilon_{\pi 2d_{3/2}} $                 & +1.4265 & +1.4259 & +2.0222 & +2.3336 & +2.3513 & +2.2581 \\
      $ \epsilon_{\pi 1h_{11/2}} $                & +1.6565 & +1.7577 & +1.6221 & +2.1503 & +2.3665 & +1.8482 \\
      $ \epsilon_{\pi 3s_{1/2}} $                 & +1.9030 & +1.8969 & +2.3310 & +2.5316 & +2.8762 & +2.9433 \\
      \hline
      $ R_p $          & 5.2099 & 5.2028 & 5.2091 & 5.2036 & 5.1928 & 5.1928 \\
      $ R_n $          & 5.0911 & 5.0834 & 5.0834 & 5.0937 & 5.0937 & 5.0922 \\
      $ R_{\urm{ch}} $ & 5.2682 & 5.2612 & 5.2675 & 5.2620 & 5.2513 & 5.2513 \\
    \end{tabular}
  \end{ruledtabular}
\end{table*}
\subsection{One-proton emission half-life}
\par
The one-proton emission half-life is estimated by the WKB approximation.
As discussed above, we assume that a $ 3s_{1/2} $ proton is emitted.
\par
The WKB approximation for one-proton emission half-life is given by~\cite{
  Buck1992Phys.Rev.C45_1688,
  Aaberg1997Phys.Rev.C56_1762,
  Xiao2023Phys.Lett.B845_138160}
\begin{equation}
  \label{eq:WKB_half_life}
  T_{1/2}^p
  =
  \frac{\hbar c \log 2}{S \Gamma}
  \, \mathrm{fm}/c,
\end{equation}
where the spectroscopic factor $ S = 2 $ is for the $ 3s_{1/2} $ protons 
and the decay width $ \Gamma $ is calculated by
\begin{equation}
  \label{eq:WKB_Gamma}
  \Gamma
  =
  N
  \frac{\left( \hbar c \right)^2}{4 \mu c^2}
  \exp
  \left[
    -
    2
    \int_{r_1}^{r_2}
    k \left( r \right)
    \, dr
  \right]
  \, \mathrm{MeV}.
\end{equation}
Here, $ \mu $ is the proton reduced mass,
$ N $ is the normalization factor given by
\begin{equation}
  \label{eq:WKB_N}
  N^{-1}
  =
  \int_0^{r_1}
  \frac{1}{k \left( r \right)}
  \cos^2
  \left[
    \int_0^r
    k \left( r' \right)
    \, dr'
    -
    \frac{\pi}{4}
  \right]
  \, dr
  \, \mathrm{fm}^2,
\end{equation}
and
\begin{equation}
  \label{eq:WKB_k}
  k \left( r \right)
  =
  \frac{\sqrt{2 \mu c^2 \left| E - V \left( r \right) \right|}}{\hbar c}
  \, \mathrm{fm}^{-1}.
\end{equation}
In this calculation,
$ \epsilon_{\pi 3s_{1/2}} $ is used for $ E $
and the proton mean-field potential shown in Fig.~\ref{fig:Pb164_pot} is used for $ V \left( r \right) $.
We use the bare proton mass $ m_p $ for $ \mu $ for simplicity.
Note that another choice for $ E $ is $ - S_p $, but it is smaller than $ \epsilon_{\pi 3s_{1/2}} $;
hence, the latter choice always gives longer lifetime.
Here, $ r_1 $ and $ r_2 $ ($ r_1 < r_2 $) are the classical turning point~\footnote{
  In Refs.~\cite{
    Buck1992Phys.Rev.C45_1688,
    Aaberg1997Phys.Rev.C56_1762,
    Xiao2023Phys.Lett.B845_138160},
  another crossing point $ r_0 < r_1 $ also appears.
  This is due to the centrifugal core,
  while we consider the proton $ 3s_{1/2} $ orbital, and hence this barrier does not exist.}
of a particle with the energy
$ E = \epsilon_{\pi 3s_{1/2}} $ in the mean-field $ V \left( r \right) $,
i.e., the crossing point of the solid and dotted lines in Fig.~\ref{fig:Pb164_pot}.
In SLy4, SLy5, and SkM*, $ r_2 $ is outside of the cutoff radius.
Therefore, we estimate $ r_2 = 81 e^2 / \epsilon_{\pi 3s_{1/2}} $~\footnote{
  We confirm that the proton mean-field potential can be well described with the Coulomb potential in $ r \gtrsim 10 \, \mathrm{fm} $.}.
\par
The estimations of the lifetime are summarized in Table~\ref{tab:Pb164_WKB}.
As can be seen, $ r_1 $ is almost always $ 7.5 \, \mathrm{fm} $,
while $ r_2 $ depends on the EDFs because they give different $ \epsilon_{\pi 3s_{1/2}} $.
Accordingly, the estimated half-lives range from the order of $ 0.1 \, \mathrm{ps} $ to $ 10 \, \mathrm{ns} $,
which is regarded as the uncertainty originates from an EDF.
Especially, if $ \epsilon_{\pi 3s_{1/2}} $ is small enough, as in the SLy4, SLy5, and SkM* cases,
it is possible to measure properties of $ \nuc{Pb}{164}{} $ before its decay by accelerator experiments.
Note that particle-emission half-life with about $ 4.5 \, \mathrm{ps} $ can be measured experimentally~\cite{
  Kohley2013Phys.Rev.Lett.110_152501}.
\begin{table*}[tb]
  \centering
  \caption{
    Proton-emission and alpha-decay half-lives, $ T_{1/2}^p $ and $ T_{1/2}^{\alpha} $, for $ \nuc{Pb}{164}{} $,
    where $ T_{1/2}^p $ and $ T_{1/2}^{\alpha} $ are estimated by using the WKB approximation
    and the phenomenological formula~\cite{
      Koura2012J.Nucl.Sci.Technol.49_816},
    respectively.
    The crossing point $ r_2 $ is given by the numerical calculation for the UNEDF series,
    while it is estimated by the Coulomb potential between a proton and the daughter nucleus for the others.
    See the main text for the detail.}
  \label{tab:Pb164_WKB}
  \begin{ruledtabular}
    \begin{tabular}{ldddddd}
      EDF & \multicolumn{1}{c}{SLy4} & \multicolumn{1}{c}{SLy5} & \multicolumn{1}{c}{SkM*} & \multicolumn{1}{c}{UNEDF0} & \multicolumn{1}{c}{UNEDF1} & \multicolumn{1}{c}{UNEDF2} \\
      \hline
      $ r_1 $ ($ \mathrm{fm} $)                                 &     7.5  &     7.5      &   7.5  &   7.5 &  7.5    &     7.4  \\
      $ r_2 $ ($ \mathrm{fm} $)                                 &    61.3  &    61.5      &  50.0  &  46.6 & 41.1    &    40.0  \\
      $ \epsilon_{\pi 3s_{1/2}} $ ($ \mathrm{MeV} $)            &    +1.9  &    +1.9      &  +2.3  &  +2.5 & +2.9    &    +2.9  \\
      $ N^{-1} $ ($ \mathrm{fm}^2 $)                            &     3.5  &     3.5      &   3.0  &   4.1 &    3.5  &     3.1  \\
      $ \Gamma $ ($ \times 10^{-13} \, \mathrm{MeV} $)          &     0.20 &     0.18     &  56    & 290   & 7000    & 13000    \\
      $ T_{1/2}^p $ ($ \mathrm{ps} $)                           & 11000    & 12000        &  41    &   7.8 &    0.33 &     0.18 \\
      \hline
      $ Q_{\alpha} $ ($ \mathrm{MeV} $)                         &     0.81 &     0.90     &   0.37 &   2.3 &    2.3  &     1.4  \\
      $ \log_{10} T_{1/2}^{\alpha} $ ($ \log_{10} \mathrm{s} $) &    98    &    90        & 170    &  36   &   37    &    63    \\
    \end{tabular}
  \end{ruledtabular}
\end{table*}
\subsection{Model analysis for the WKB approximation}
\par
\begin{figure}[tb]
  \centering
  \includegraphics[width=1.0\linewidth]{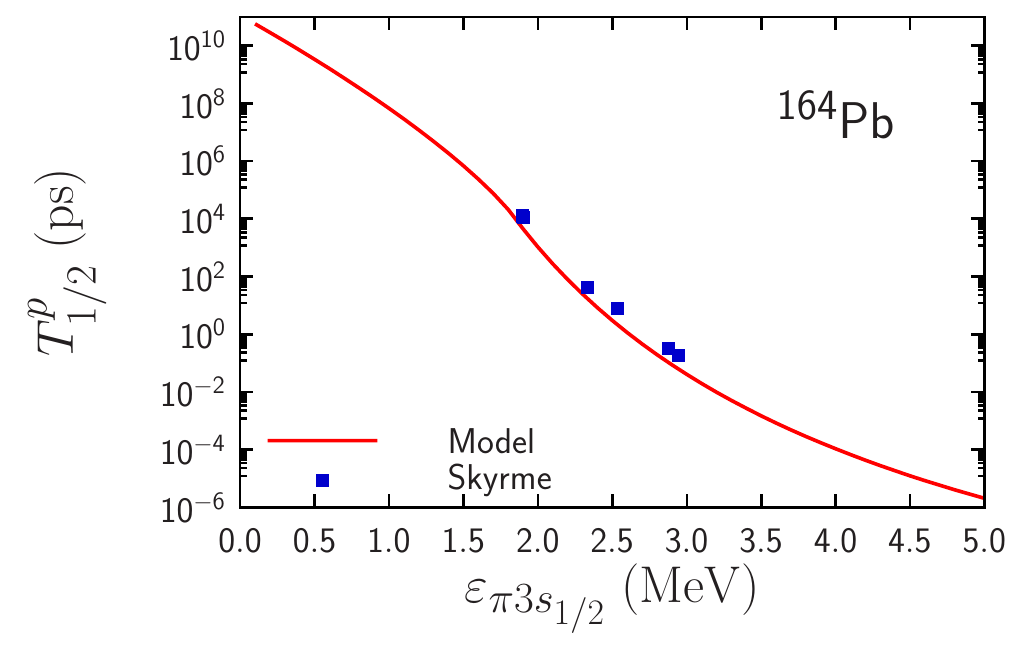}
  \caption{
    One-proton emission half-life $ T_{1/2}^p $ calculated with the model potential (solid line) and the WKB approximation (squares).
    See the text for the detail.}
  \label{fig:model_T12}
\end{figure}

To elucidate the relation between the $ 1p $-proton emission half-life and the single-particle energy of the proton $ 3s_{1/2} $ orbital,
we approximate the proton mean field by a simplified potential given by 
\begin{equation}
  V \left( r \right)
  =
  \begin{cases}
    -40 \, \mathrm{MeV}
    & \text{for $ r < 7.5 \, \mathrm{fm} $}, \\
    81 e^2 / r
    & \text{otherwise},
  \end{cases}
\end{equation}
where the classical turning points are set as
$ r_1 = 7.5 \, \mathrm{fm} $ and $ r_2 = 81 e^2 / \epsilon_{\pi 3s_{1/2}} $.
By the WKB approximation with this potential,
we estimated the half-life as a function of $ \epsilon_{\pi 3s_{1/2}} $ as shown in Fig.~\ref{fig:model_T12}.
It reasonably explains the estimation from more realistic HFB potential indicating that
the half-life is not sensitive to the detail of the potential.
Thus, as expected, the half-life is roughly determined by only the single-particle energy of the proton $ 3s_{1/2} $ orbital.
In turn, this means that the measurement of the half-life will provide an estimation of the proton single-particle energy $ \epsilon_{\pi 3s_{1/2}} $.
\subsection{Alpha-decay half-life}
\par
Lastly, we estimate the $ \alpha $-decay half-life $ T_{1/2}^{\alpha} $.
We use a phenomenological formula proposed in Ref.~\cite{
  Koura2012J.Nucl.Sci.Technol.49_816}
\begin{widetext}
  \begin{equation}
    \label{eq:alpha_half_life}
    \log_{10} T_{1/2}^{\alpha}
    =
    \frac{2 \log_{10} e}{\hbar c}
    \sqrt{2 M_{\alpha} \frac{A - 4}{A}}
    \int_R^b
    \sqrt{V_{\urm{C}} \left( r \right) - Q_{\alpha}}
    \, dr
    -
    \log_{10} N
    -
    \log_{10} \left( \log_2 e \right)
    \, 
    \log_{10} \mathrm{s}
  \end{equation}
\end{widetext}
with
\begin{subequations}
  \begin{align}
    V_{\urm{C}}  \left( r \right)
    & =
      \frac{2 Z_{\urm{D}} e^2}{r}, \\
    b
    & =
      \frac{2 Z_{\urm{D}} e^2}{Q_{\alpha}}
  \end{align}
\end{subequations}
where $ A = 164 $ is the mass number of the parent nucleus,
$ Z_{\urm{D}} = 80 $ is the proton number of the daughter nucleus,
and 
$ M_{\alpha} $ is the mass of the $ \alpha $ particle.
The parameter $ - \log_{10} N = -21.4577 $ is related to the penetration factor and the preformation probability;
$ R $ is the position where the penetration of the potential barrier begins
and
$ R = r_0 A_{\urm{D}}^{1/3} + d_0 $,
$ r_0 = 1.08 \, \mathrm{fm} $, and $ d_0 = 2.0 \, \mathrm{fm} $
are used in this calculation
with the mass number of the daughter nucleus $ A_{\urm{D}} = 160 $.
The estimated values of $ T_{1/2}^{\alpha} $ are shown in Table~\ref{tab:Pb164_WKB}.
It can be seen that $ T_{1/2}^{\alpha} $ is more than $ 10^{35} \, \mathrm{s} $;
thus, we need not to consider the $ \alpha $ decay of $ \nuc{Pb}{164}{} $.
Note that even if the deformation is considered, $ Q_{\alpha} $ does not change much and hence,
the conclusion does not change.
%
\section{Summary}
\par
Using the Skyrme Hartree-Fock-Bogoliubov calculation,
we confirmed that a $ \nuc{Pb}{164}{} $ nucleus is a possible heaviest $ N = Z $ doubly magic nucleus.
The proton-emission half-life is estimated by using the Wentzel-Kramers-Brillouin approximation.
The estimated half-life ranges from $ 0.1 \, \mathrm{ps} $ to $ 10 \, \mathrm{ns} $,
which depends on the single-particle energy of the proton $ 3s_{1/2} $ orbital.
There is a one-to-one correspondence between the proton single-particle energy of the $ 3s_{1/2} $ orbital
and the proton-emission half-life.
If the single-particle energy is lower, the half-life is longer and it may be possible to investigate properties of $ \nuc{Pb}{164}{} $  experimentally.
In turn, once the proton-emission half-life is measured, the proton single-particle energy can be estimated,
which will impose a constraint on energy density functionals.
The alpha-decay half-life is estimated to be longer than $ 10^{35} \, \mathrm{s} $,
which is much longer than the age of the universe;
hence, the alpha decay of $ \nuc{Pb}{164}{} $ can be safely neglected although the alpha-decay channel is opened.
The recent experiment~\cite{
  Yang2024Phys.Rev.Lett.132_072502}
predicted that the $ N = 82 $ magic number may be still robust in the neutron deficient side.
Our finding also supports this prediction,
and the synthesis of $ \nuc{Pb}{164}{} $ at next-generation accelerator facilities is demanded.
\appendix
\section{Calculation results for $ \nuc{Sn}{100}{} $}
\label{sec:appendix_Sn100}
\par
The calculation results for $ \nuc{Sn}{100}{} $ are shown in Table~\ref{tab:Sn100_result}.
The calculation setup is the same as for $ \nuc{Pb}{164}{} $ shown in the main text.
Since the experimental data of masses are available for $ \nuc{Cd}{98}{} $ and $ \nuc{Sn}{100}{} $ in the AME2020~\cite{
  Huang2021Chin.Phys.C45_030002,
  Wang2021Chin.Phys.C45_030003},
the experimental $ S_{2p} $ is also shown.
Although the experimental data of masses of $ \nuc{Cd}{96}{} $ and $ \nuc{In}{99}{} $ are not available,
estimated values for them are available in the AME2020.
Hence, the estimated values for $ S_p $ and $ Q_{\alpha} $ are also shown for the reference.
\begin{table*}[tb]
  \centering
  \caption{
    Same as Table I but for $ \nuc{Sn}{100}{} $.
    The experimental data taken from Refs.~\cite{
      Huang2021Chin.Phys.C45_030002,
      Wang2021Chin.Phys.C45_030003}
    are also shown,
    where $ E \left( \nuc{Cd}{96}{} \right)  $ and $ E \left( \nuc{In}{99}{} \right)  $ are the estimated values.}
  \label{tab:Sn100_result}
  \begin{ruledtabular}
    \begin{tabular}{lddddddd}
      EDF & \multicolumn{1}{c}{SLy4} & \multicolumn{1}{c}{SLy5} & \multicolumn{1}{c}{SkM*} & \multicolumn{1}{c}{UNEDF0} & \multicolumn{1}{c}{UNEDF1} & \multicolumn{1}{c}{UNEDF2} & \multicolumn{1}{c}{Expt.} \\
      \hline
      $ E \left( \nuc{Sn}{100}{} \right) $        & -828.5887 & -827.8619 & -826.1837 & -826.8630 & -828.8568 & -829.7869 & -825.16260 \\
      $ E \left( \nuc{Cd}{98}{} \right)  $        & -823.8963 & -823.2002 & -820.9489 & -822.4334 & -824.6167 & -824.8113 & -821.07294 \\
      $ E \left( \nuc{Cd}{96}{} \right)  $        & -794.0282 & -793.5384 & -790.4066 & -793.3220 & -795.1982 & -794.4453 & -792.9     \\
      $ E \left( \nuc{In}{99}{} \right)  $        & -825.4207 & -824.7584 & -822.9540 & -824.0293 & -826.0985 & -826.5907 & -822.1     \\
      \hline
      $ S_p $                                     &  3.1680 & 3.1036 & 3.2297 & 2.8337 & 2.7584 & 3.1962 &  3.1     \\
      $ S_{2p} $                                  &  4.6924 & 4.6617 & 5.2348 & 4.4296 & 4.2401 & 4.9756 &  4.08966 \\
      $ Q_{\alpha} $                              & -6.2648 & -6.0279 & -7.4814 & -5.2454 & -5.3630 & -7.0459 & -4.0     \\
      \hline
      $ \Delta_p \left( \nuc{Sn}{100}{} \right) $ & 0.0007 & 0.0006 & 0.0005 & 0.0006 & 0.0006 & 0.0006 & \\
      $ \Delta_n \left( \nuc{Sn}{100}{} \right) $ & 0.0005 & 0.0004 & 0.0008 & 0.0019 & 0.0006 & 0.0003 & \\
      $ \Delta_p \left( \nuc{Cd}{98}{} \right)  $ & 0.8218 & 0.7727 & 0.6123 & 0.6189 & 0.6383 & 0.7084 & \\
      $ \Delta_n \left( \nuc{Cd}{98}{} \right)  $ & 0.0022 & 0.0019 & 0.0012 & 0.0038 & 0.0025 & 0.0017 & \\
      \hline
      $ \epsilon_{\pi 1f_{5/2}} $                 & -7.9892 & -8.1391 & -6.6304 & -4.8840 & -5.9802 & -5.8387 & \\
      $ \epsilon_{\pi 2p_{3/2}} $                 & -7.3229 & -7.3311 & -6.6013 & -5.7452 & -5.6142 & -5.7212 & \\
      $ \epsilon_{\pi 2p_{1/2}} $                 & -5.6751 & -5.6918 & -4.9295 & -4.3633 & -4.3989 & -4.5109 & \\
      \hline
      $ R_p $          & 4.4287 & 4.4231 & 4.4210 & 4.4223 & 4.4076 & 4.4069 \\
      $ R_n $          & 4.3478 & 4.3418 & 4.3365 & 4.3504 & 4.3440 & 4.3420 \\
      $ R_{\urm{ch}} $ & 4.4974 & 4.4918 & 4.4897 & 4.4911 & 4.4766 & 4.4759 \\
    \end{tabular}
  \end{ruledtabular}
\end{table*}
\section{Calculation results for $ N = Z $ nuclei using HFBTHO code}
\label{sec:appendix_HFBTHO}
\par
To show the $ \nuc{Pb}{164}{} $ nucleus shows typical behaviur of a doubly magic nucleus,
we calculate the $ N = Z $ even-even nuclei ($ 8 \le N = Z \le 100 $) with considering the axial deformation.
We use the HFBTHO code~\cite{
  NavarroPerez2017Comput.Phys.Commun.220_363}.
Harmonic-oscillator wave functions up to $ 24 \hbar \omega $ are used for the basis
with $ \hbar \omega = 1.2 \times 41 A^{-1/3} \, \mathrm{MeV} $.
In order to consider the deformation properly,
each nucleus is calculated with the initial Woods-Saxon potential of different deformation parameter
$ \beta_2 = -0.5 $, $ -0.4 $, \ldots, $ +0.5 $,
and the result with the smallest energy is used for the further analysis.
We take the SLy4 EDF with the volume-type pairing interaction as an example.
\par
\begin{figure}[tb]
  \centering
  \includegraphics[width=1.0\linewidth]{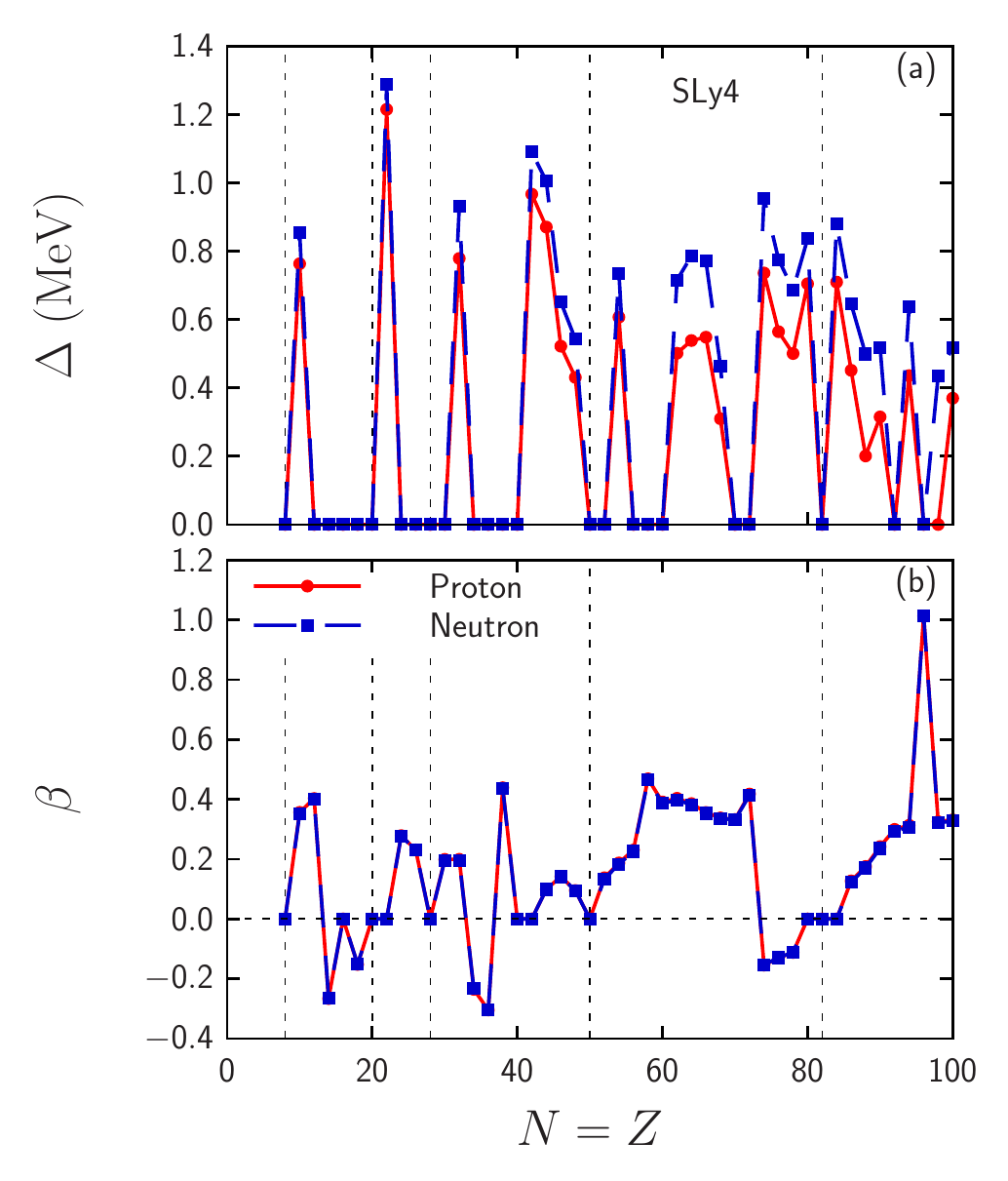}
  \caption{
    Pairing gap $ \Delta $ and deformation parameter $ \beta_2 $ for protons (solid line) and neutrons (dashed line)
    calculated by using the HFBTHO code with the SLy4 EDF and the volume-type pairing interaction.}
  \label{fig:HFBTHO_1}
\end{figure}

Figure~\ref{fig:HFBTHO_1} shows the pairing gap $ \Delta $ and the deformation parameter $ \beta_2 $ for protons and neutrons.
As it can be seen, the $ N = Z $ nuclei with the conventional magic numbers, $ Z $ or $ N = 8 $, $ 20 $, $ 28 $, $ 50 $, and even $ 82 $,
are spherical ($ \beta_2 = 0 $) without the pairing gap ($ \Delta = 0 $).
The deformation parameter $ \beta_2 $ for protons are always almost the same as those for neutrons.
\par
\begin{figure}[tb]
  \centering
  \includegraphics[width=1.0\linewidth]{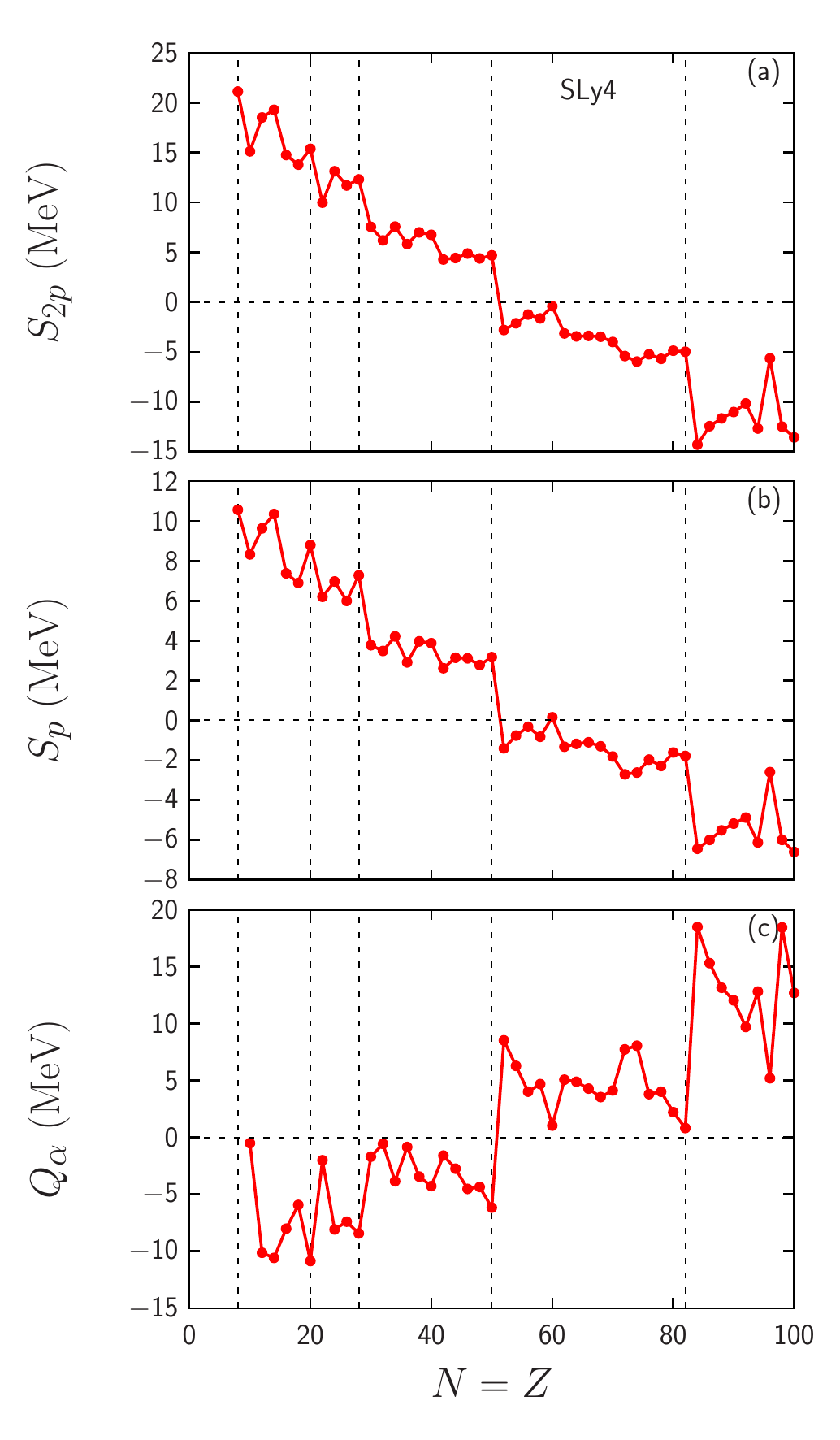}
  \caption{
    Two-proton separation energy $ S_{2p} $,
    one-proton separation energy $ S_p $,
    and
    the $ \alpha $-decay $ Q $-value $ Q_{\alpha} $ 
    calculated by using the HFBTHO code with the SLy4 EDF and the volume-type pairing interaction.}
  \label{fig:HFBTHO_2}
\end{figure}

Figure~\ref{fig:HFBTHO_2} shows the two-proton separation energy $ S_{2p} $,
the one-proton separation energy $ S_p $,
and the $ \alpha $-decay $ Q $-value $ Q_{\alpha} $.
Energies of odd-$ A $ nuclei are needed to calculate $ S_p $;
we take the similar way to Eq.~\eqref{eq:Sp_approx}.
However, in contrast to $ \nuc{Pb}{164}{} $,
there is an ambiguity or the approximation of $ \Delta_p $ of odd-$ A $ nuclei.
Therefore, we calculate the energy of an odd-$ A $ nucleus by
\begin{equation}
  \label{eq:Sp_general}
  E \left( Z - 1, N \right)
  \simeq
  \frac{E \left( Z - 2, N \right) + E \left( Z, N \right)}{2}
  +
  \overline{\Delta}_p
\end{equation}
with 
\begin{equation}
  \overline{\Delta}_p
  =
  \begin{cases}
    \Delta_p \left( Z, N \right)
    & \text{if $ Z - 2 $ is magic}, \\
    \Delta_p \left( Z - 2, N \right)
    & \text{if $ Z $ is magic}, \\
    \frac{\Delta_p \left( Z - 2, N \right) + \Delta_p \left( Z, N \right)}{2}
    & \text{otherwise},
  \end{cases}
\end{equation}
where the magic numbers used here are conventional $ 8 $, $ 20 $, $ 28 $, $ 50 $, and $ 82 $.
The magic numbers, including $ Z = 82 $, show sudden change of $ S_{2p} $, $ S_p $, and $ Q_{\alpha} $.
Therefore, $ N = Z = 82 $ behaves as the normal magic numbers.
%
%
\begin{acknowledgments}
  The authors acknowledge the fruitful discussion with
  Shintaro Go,
  Hiroyuki Koura,
  Hiroyuki Sagawa,
  Hiroyoshi Sakurai,
  Yutaka Shikano, 
  Daisuke Suzuki,
  and the members of the RIBF Future Plan Working Group.
  T.N.~acknowledges
  the RIKEN Special Postdoctoral Researcher Program,
  the JSPS Grant-in-Aid for Research Activity Start-up under Grant No.~JP22K20372,
  the JSPS Grant-in-Aid for Transformative Research Areas (A) under Grant No.~JP23H04526,
  the JSPS Grant-in-Aid for Scientific Research (B) under Grant No.~JP23H01845 and No.~JP23K01845,
  the JSPS Grant-in-Aid for Scientific Research (C) under Grant No.~JP23K03426,
  and 
  the JSPS Grant-in-Aid for Early-Career Scientists under Grant No.~JP24K17057.
  The numerical calculations were performed on cluster computers at the RIKEN iTHEMS program.
  This work was supported by the RIKEN TRIP initiative (Nuclear transmutation).
\end{acknowledgments}
%
%
%
\end{document}